\documentclass[pr,twocolumn,showpacs,aps]{revtex4}
\usepackage{amsfonts,amssymb,amsmath,bm}
\usepackage{wrapfig}
\usepackage[dvips]{graphicx}
\usepackage[dvips]{color}
\usepackage[dvipsnames]{xcolor}
\usepackage{times}
\usepackage{marvosym}
\begin{document}

\title{Adjustable subwavelength localization in a hybrid plasmonic waveguide}
\author{S. Belan$^{1,2}$, S. Vergeles$^{1,2}$, P. Vorobev$^{1,2}$}
 \affiliation{$^1$Landau Institute for Theoretical Physics RAS, Kosygina 2, 119334,
 Moscow, Russia}
 \affiliation{$^2$Moscow Institute of Physics and Technology, Dolgoprudnyj, Institutskij
 per. 9, 141700, Moscow Region, Russia}



\begin{abstract}
The hybrid plasmonic waveguide consists of a high-permittivity dielectric nanofiber embedded in a low-permittivity dielectric near a metal surface. This architecture is considered as one of the most perspective candidates for long-range subwavelength guiding. We present qualitative analysis and numerical results which reveal advantages of the special waveguide design when dielectric constant of the cylinder is greater than the absolute value of the dielectric constant of the metal. In this case the arbitrary subwavelength mode size can be achieved by controlling the gap width. Our qualitative analysis is based on consideration of sandwich-like conductor-gap-dielectric system. The numerical solution is obtained by expansion of the hybrid plasmonic mode over single cylinder modes and the surface plasmon-polariton modes of the metal screen and matching the boundary conditions.
\end{abstract}


\maketitle

\section{Introduction}
The creation of the waveguides capable of guiding light with deep
subwavelength confinement is of great interest for practical
applications. These devices may throw open the doors to nanoscale
optical communications, quantum computing, nanoscale lasers and
bio-medical sensing. The main problem on the way to practical
realization is the diffraction limit of light in dielectric media.
Electromagnetic energy cannot be localized into nanoscale region
much smaller than the wavelength of light in the dielectric
\cite{Born_1999_book}. The possible solution to this problem is using of the
materials with negative dielectric permittivity.
For example,  metals are known to exhibit this property below the plasma frequency.
Metal structures provide guiding of the surface plasmon-polaritons (SPP),
which can be strongly localized near metal-dielectric interfaces\cite{Gramotnev_2010_NaturePhot}.
However the propagation length of the strongly confined plasmonic modes is not large enough due to the presence of Ohmic losses
in the dissipative metal regions.

The new approach for this challenge integrates dielectric waveguide
with plasmonic one. The hybrid plasmonic waveguide consists of a
high-permittivity dielectric nanofiber separated from a metal
screen by low-permittivity dielectric nanoscale gap \cite{XiangZhang_2008_NaturePhot}.
Both the single fiber and the silver-dielectric interface cannot provide strong mode confinement at optical and near infrared frequencies,
but such hybrid conductor-gap-dielectric architecture has experimentally
demonstrated deep subwavelength optical waveguiding \cite{XiangZhang_2011_NatureComm}.
Relatively large propagation distance has been achieved due to
low loss tangent $tg=\varepsilon_{\mathrm{m}}^{\prime\prime}/\varepsilon_{\mathrm{m}}^\prime$
at the operating frequency and the specific spatial structure of the guiding mode with field confinement within non dissipative gap region.

In the present paper we show that the hybrid plasmon polariton(HPP) mode confinement can be considerably risen by a specific choice of the materials,
when the dielectric constant of the cylinder is greater than the absolute
value of the dielectric constant of the metal screen.
The main advantage of the choice is the hyperbolic-like dependence
of the effective index on the gap width.
This feature allows to achieve arbitrary subwavelength mode size
at any frequency by tuning the distance between the cylinder and the metal.
To justify our approach we theoretically investigate propagation of the HPP-mode.
First we give qualitative analysis 
basing on the consideration of plane sandwich-like
conductor-gap-dielectric waveguide structure(CGD)
\cite{Avrutsky_2010_OptExpress}.
We derive exact analytical expression for effective index of the fundamental CGD-mode
and give criterion when the HPP-mode is CGD-like.
Finally, we present semi-analytical approach for describing of the HPP-mode propagation.
Similar approach has been previously applied for plane wave scattering by a cylinder placed near the plane surface \cite{Frezza_1996}-\cite{Frezza_1997}.

The scheme is based on expansion of the HPP-mode over single
cylinder modes and the surface plasmon polariton modes of the metal-dielectric interface and matching the boundary conditions for electromagnetic field components.
Numerically obtained dispersion relations confirm the advantages
of our design of the hybrid waveguide.
\begin{figure}[h]
 \center{\includegraphics[scale=0.75]{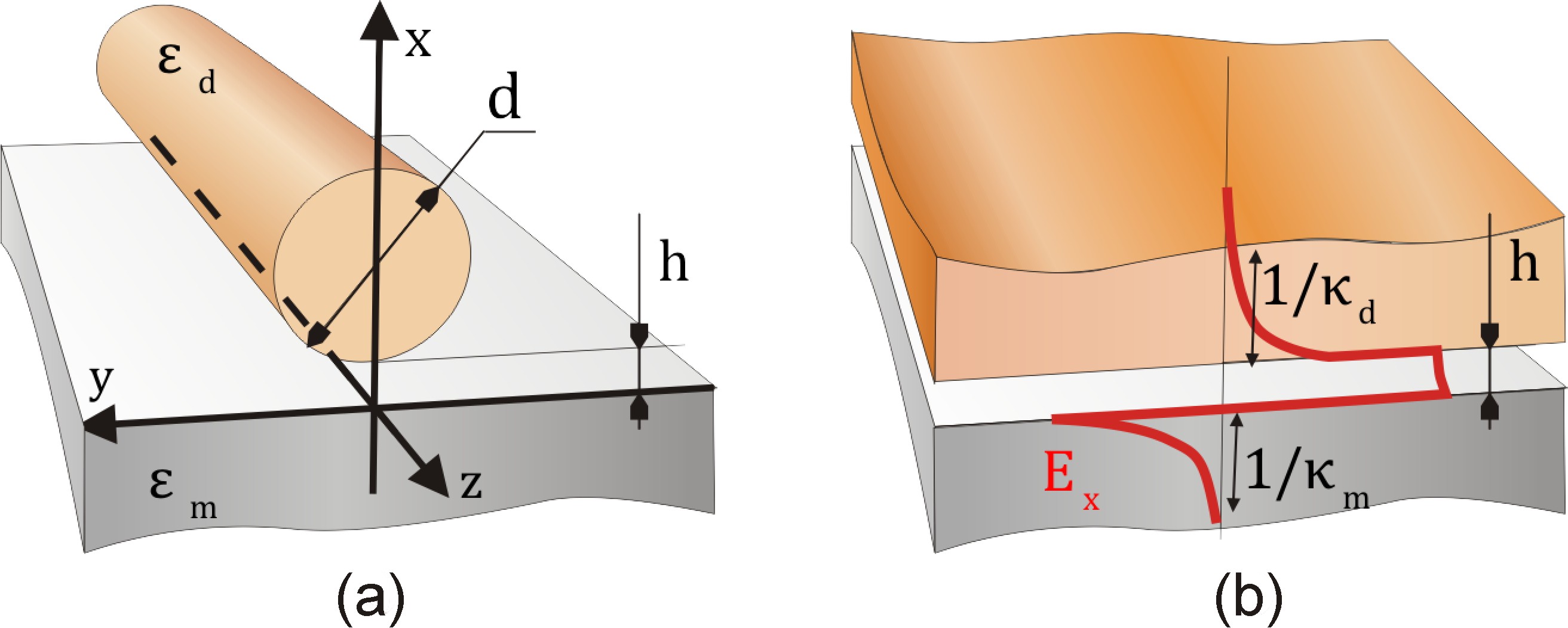}}
 \caption{  {\bf a)} Geometry of the waveguide; \ \ \ {\bf b)}
  Plain waveguide with the same width of the gap.}
 \label{pic:3D_view}
\end{figure}

\section{Qualitative description}
\label{section:qualitative}
The geometry of the hybrid waveguide is the following: a circular
dielectric cylinder of diameter $d$ and permittivity $\varepsilon_d$ is placed above a
metal screen of permittivity $\varepsilon_m$. The 
width of the gap between the cylinder and the metal screen is $h$,
see Fig.~\ref{pic:3D_view}$(a)$.
Let us choose the Cartesian reference system as it is shown in
Fig.~\ref{pic:3D_view}: $z$-axis is directed along the waveguide,
whereas $x$-axis is directed normally to the metal screen.
We consider a plasmon-polariton mode of frequency $\omega$ and the propagation constant $\beta$ propagating along $z$-axis. 
Thus, all electromagnetic field components depend on time and
$z$-coordinate as $\exp[i\beta z - i \omega t]$. We assume, that
responses of both dielectric and metal on electromagnetic field are
described by dielectric constants, which are
$\varepsilon_{\mathrm{m}}$ and $\varepsilon_{\mathrm{d}}$
respectively. Generally, the outer medium may be not vacuum, but
some dielectric medium having dielectric constant being equal to
$\varepsilon_\mathrm{g}$. All the materials are assumed to be
nonmagnetic. To describe the mode confinement, it is convenient to introduce
effective refractive index $n_{\mathrm{eff}}$, which is
defined as $n_{\mathrm{eff}}=\beta/k$, where $k=\omega/c$ in the
wavenumber in vacuum. The effective index determines the field penetration
depth into the material with permittivity $\varepsilon$ as
$1/k\sqrt{n_{\mathrm{\mathrm{eff}}}^2 - \varepsilon}$.
The penetration depth of the bound mode should be real in the unbounded waveguide constituents
(metal and outer dielectric space),
and may be imaginary for bounded constituents (fiber).
The greater $n_{\mathrm{eff}}$ is the stronger degree of confinement.

Optimization for transversal field confinement implemented in paper
\cite{XiangZhang_2008_NaturePhot} for hybrid waveguide shows that
the thinner gaps provide higher localization. The fiber diameter is
much greater than the the gap width in the case,
and the mode is sufficiently localized in the region where the gap can be considered as
approximately plain. In the region, the waveguide shape is close
to plain sandwich like conductor-gap-dielectric (CGD) structure, see Fig.~\ref{pic:3D_view}$(b)$. The
limit of plain CGD-model was considered in
\cite{Avrutsky_2010_OptExpress}, where the properties of the bound
fundamental mode were investigated. One of the main advantage of the
CGD-mode is that effective index of the mode $n_{\mathrm{CGD}}$ is
greater than the refractive index of the dielectric
$\sqrt{\varepsilon_{\mathrm{d}}}$,
$n_{\mathrm{CGD}}>\sqrt{\varepsilon_{\mathrm{d}}}$. This implies, that the
electromagnetic field of the mode decays exponentially into the
dielectric cladding. Nevertheless, the analysis proposed in
\cite{Avrutsky_2010_OptExpress} is not applicable to HPP-mode of
hybrid waveguide with optimal diameter found in
\cite{XiangZhang_2008_NaturePhot}. The reason is that the mode of
plain CGD-model indeed describes the HPP-mode
only for large enough fiber diameter $d$ otherwise HPP-mode should
be considered as a result of hybridization of surface plasmon
polariton modes and the modes of the single dielectric cylinder.
\begin{equation}\label{h_c}
	h_c=
	\frac
	{\lambda}
	{4\pi\sqrt{\varepsilon_{\mathrm{d}}-\varepsilon_\mathrm{g}}}
	\log\frac{\varepsilon_{\mathrm{m}}\sqrt{\varepsilon_{\mathrm{d}}
	-\varepsilon_\mathrm{g}}-\varepsilon_\mathrm{g}\sqrt{\varepsilon_{\mathrm{d}}
	-\varepsilon_{\mathrm{m}}}}
	{\varepsilon_{\mathrm{m}}\sqrt{\varepsilon_{\mathrm{d}}
-\varepsilon_\mathrm{g}}+\varepsilon_\mathrm{g}\sqrt{\varepsilon_{\mathrm{d}}
-\varepsilon_{\mathrm{m}}}}.
\end{equation}
The main goal of the present work is
to give the theoretical description of the hybrid waveguide and to find approaches to deeper
localization of the HPP-mode. Comparative
analysis of
\cite{XiangZhang_2008_NaturePhot,Avrutsky_2010_OptExpress} suggests,
that in order to get stronger transversal miniaturization of the
hybrid waveguide the CGD-like regime of propagation
(with $n_{\mathrm{eff}} > \sqrt{\varepsilon_{\mathrm{d}}}$) should be
achieved for the diameter which is much less than free space
wavelength. Our analysis of the plain CGD-structure shows that the
localization of the fundamental mode can be significantly risen for
special set of materials, when absolute value of the metal
dielectric constant is less than the dielectric constant of the
dielectric cladding,
$|\varepsilon_{\mathrm{m}}|<\varepsilon_{\mathrm{d}}$.
For the case, the effective refractive index
$n_{\scriptscriptstyle \mathrm{CGD}}$ is proportional to
inverse width of the gap, $n_{\scriptscriptstyle \mathrm{CGD}}\propto 1/kh$, when the
width $h$ is small enough. To use the same effect for the hybrid
waveguide, the cylinder diameter should sufficiently exceed some critical value
$d^\ast$, which is determined by the condition that the transversal
size of the plain part of the gap is comparable with the mode
penetration depth into the dielectric $1/k\sqrt{n_{\mathrm{\mathrm{eff}}}^2 - \varepsilon_d}$. The
size of the plain part of the gap is evaluated as
2$\sqrt{hd}$, thus the condition is $2\sqrt{hd^\ast}
\approx 1/k\sqrt{n_{\mathrm{\mathrm{eff}}}^2 - \varepsilon_d}$. For $d$ greater enough than $d^\ast$ the
guiding mode can approach the strongly confined mode of the sandwich
like system even if the diameter of the cylinder is much less than
free space wavelength.

In order to give general physical argumentation of our results let
us consider planar sandwich-like CGD-waveguide in detail.
The wave vector of fundamental CGD-mode (which is TM-mode) can be
calculated from equation~\cite{Avrutsky_2010_OptExpress}
\begin{equation}
	\exp[2h\kappa_\mathrm{g}]
	=
	\frac{
	(
	\varepsilon_{\mathrm{d}}\kappa_\mathrm{g}
	-
	\varepsilon_\mathrm{g}\kappa_{\mathrm{d}}
	)
	(
	\varepsilon_{\mathrm{m}}\kappa_\mathrm{g}
	-
	\varepsilon_\mathrm{g}\kappa_{\mathrm{m}}
	)
	}
	{
	(
	\varepsilon_{\mathrm{d}}\kappa_\mathrm{g}
	+
	\varepsilon_\mathrm{g}\kappa_{\mathrm{d}}
	)
	(
	\varepsilon_{\mathrm{m}}\kappa_\mathrm{g}
	+
	\varepsilon_\mathrm{g}\kappa_{\mathrm{m}}
	)
	},
\end{equation}
where $\kappa_{i} = k\sqrt{n_{\mathrm{\scriptscriptstyle
\mathrm{CGD}}}^2 - \varepsilon_{i}}$ for each material,
$i= m,g,d$ and
$n_{\scriptscriptstyle \mathrm{CGD}}$ is the effective index of the
mode. In particular, $1/\kappa_{\mathrm{d}}$ and
$1/\kappa_{\mathrm{m}}$ are the penetration depths into the
dielectric and the metal correspondingly. It is known that such
plane three-layer waveguide supports the propagation of the bound
eigen mode only if the width of the intermediate layer is less than
some cut-off value $h_c$ which is determined by the permittivities
at given frequency

When the thickness exceeds this critical value the fundamental CGD-mode
becomes radiative with energy leaking into upper dielectric half
space.

\begin{figure}[t]
 \center{\includegraphics[scale=0.4]{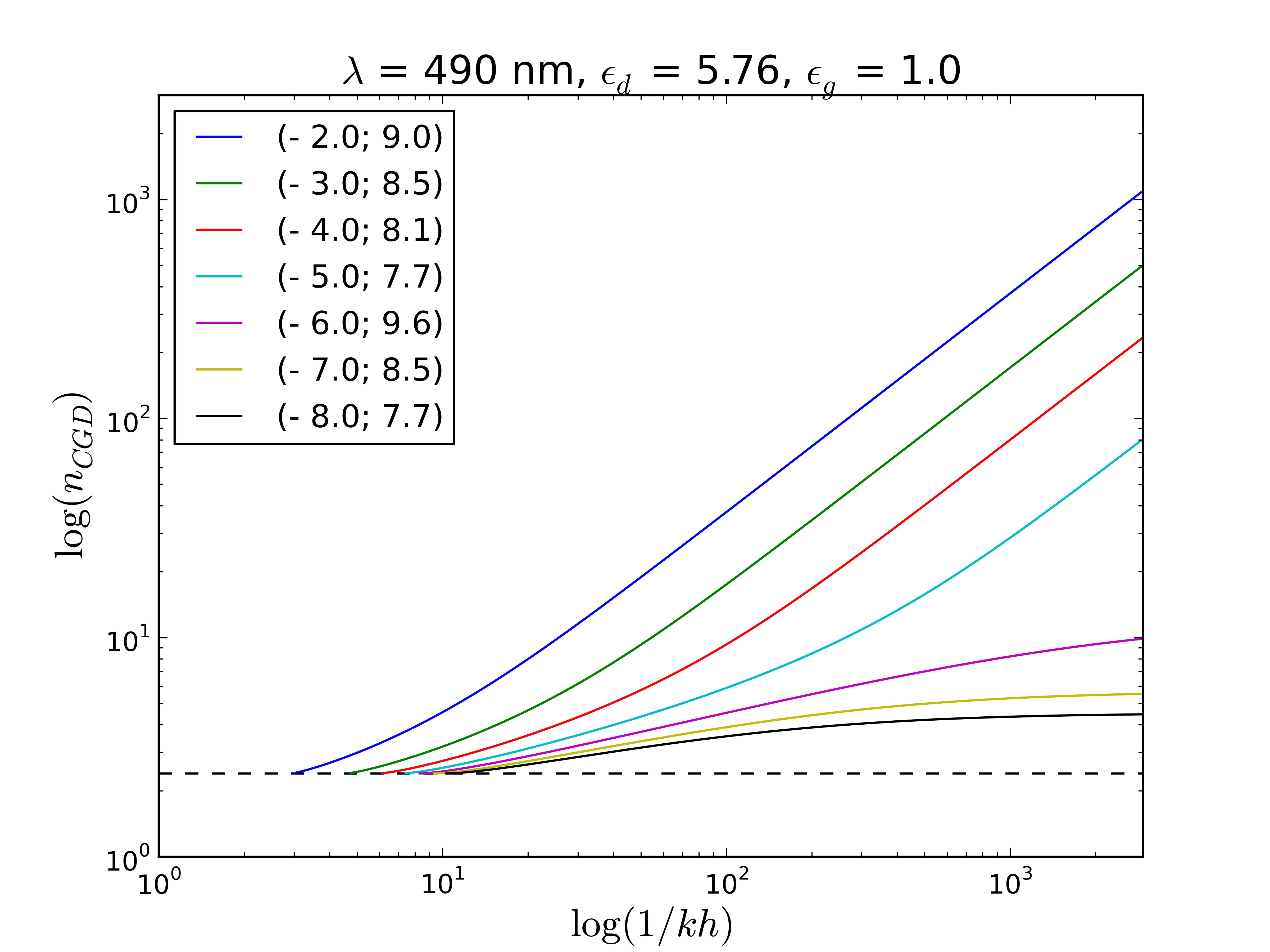}}
 \caption{(Color online) Effective index $n_{\scriptscriptstyle\mathrm{CGD}}$ of the CGD-mode versus gap width $h$ for different metal permittivity and critical gap width $(\varepsilon_m; h_c)$.
 The dielectric constants of the dielectric and gap region are
 $\varepsilon_\mathrm{d}=5.76$ and $\varepsilon_\mathrm{g}=1$ respectively at wavelength $\lambda = 490nm$.}
 \label{pic:CGD}
\end{figure}

There is a significant difference between the dependence of the
effective index $n_{\scriptscriptstyle\mathrm{CGD}}$ on the gap
thickness $h$ for the cases of low and high index dielectric, see Fig.\ref{pic:CGD}. For
relatively low refractive index of dielectric,
$\varepsilon_\mathrm{d} < |\varepsilon_\mathrm{m}|$, there exists
surface plasmon-polariton mode when the gap is absent, $h=0$. It has
effective index $n_{\mathrm{md}} =
\sqrt{\varepsilon_{\mathrm{m}}\varepsilon_{\mathrm{d}}
/(\varepsilon_{\mathrm{d}}+\varepsilon_{\mathrm{m}})}$. Then the
effective index of the fundamental CGD-mode is bounded,
$\sqrt{\varepsilon_{\mathrm{d}}} < n_{\scriptscriptstyle\mathrm{CGD}} <
n_{\mathrm{md}}$. Just the case was considered in the papers
\cite{Avrutsky_2010_OptExpress} and
\cite{XiangZhang_2008_NaturePhot}. Otherwise when permittivity of dielectric
is relatively high
\begin{equation}\label{our_scheme}
\varepsilon_\mathrm{d} > |\varepsilon_\mathrm{m}| > \varepsilon_\mathrm{g},
\end{equation}
the effective index
$n_{\scriptscriptstyle\mathrm{CGD}}$ unlimitedly diverges as the gap
thickness tends to zero, $h\ll \lambda/\sqrt{\varepsilon_{\mathrm{d}}}$:
\begin{equation}\label{n_CGD}
	n_{{\scriptscriptstyle\mathrm{CGD}}}
	\approx
	\frac{1}{2kh}\ln\frac
	{(\varepsilon_{\mathrm{d}}-\varepsilon_\mathrm{g})
	 (\varepsilon_{\mathrm{m}}-\varepsilon_\mathrm{g})}
	{(\varepsilon_{\mathrm{d}}+\varepsilon_\mathrm{g})
	 (\varepsilon_{\mathrm{m}}+\varepsilon_\mathrm{g})}.
\end{equation}
This leads to extremely strong
light confinement in a transparent dielectric gap layer located
between the high-index dielectric and the conductor. The actual
degree of localization is restricted only by additional factors,
such as increasing Ohmic losses in the metal, spatial dispersion and
atomic structure of the materials. In this respect the properties of the conductor-gap-dielectric plasmonic mode similar to that of the gap plasmon polaritons in a conductor-gap-conductor structure \cite{Tanaka_2003}-\cite{Avrutsky_2007}. This feature is the principle
behind our idea: in practice one should choose the metal of the
absolute permittivity less than the permittivity of the cylinder and
place cylinder at distance $h < h_\mathrm{c}$ from the metal plane.
When such metal is involved the effective index of the HPP mode can
be significantly greater than effective index of electromagnetic
field in bulk material of the cylinder even for very small diameters of the cylinder.
Note, that to calculate group velocity $v_g$ and chromatic
dispersion for the mode using the formula (\ref{n_CGD}),
one should know the dispersion laws for permittivities
$\varepsilon_{\mathrm{m}}$ and $\varepsilon_{\mathrm{d}}$.
For thin gap $h\ll h_c$, the group velocity scales as $v_g/c \propto h/\lambda$.
Thus divergence of the CGD-mode effective index with gap width decreasing leads to strong reduction of the group velocity.

There is reverse side of the strong localization which is small propagation distance.
It was shown in paper \cite{XiangZhang_2008_NaturePhot} that the strongest localization of the HPP-mode
corresponds to the lowest propagation length.
It is common place of waveguides which use metal as a constructive component.
Let us consider limit when the gap index is low, so
$\varepsilon_\mathrm{g}\ll |\varepsilon_{\mathrm{m}}|$.
For the case
\begin{equation}\label{n_CGD_losses}
	n_{\scriptscriptstyle\mathrm{CGD}}
	\approx
	\frac{1}{kh|\varepsilon_{\mathrm{m}}|}
	\left(
	1 -
	\frac{|\varepsilon_{\mathrm{m}}|}{\varepsilon_{\mathrm{d}}}
	+
	i \frac{\varepsilon_{\mathrm{m}}^{\prime\prime}}{|\varepsilon_{\mathrm{m}}|}
	\right),
\end{equation}
where $\varepsilon_{\mathrm{m}}^{\prime\prime}$ is the imaginary part of the metal
permittivity.
It follows from Eq. (\ref{n_CGD_losses}), that the
localization radius is of the order of $h|\varepsilon_m|$
in the limit $h\ll h_c$.
Note, that our approach allows to
squeeze the mode at arbitrary frequency into any subwavelength scale
simply by tuning the gap width in accordance with (\ref{n_CGD}).
Hence, our waveguide design breaks connection between mode localization and
the carrying frequency of the mode.
In particular, the approach may be interesting for design waveguides at THz frequencies
\cite{XZhang_2009_PRL,Nam_2009_OptExp}.
The propagation length
$\ell \sim h|\varepsilon_{\mathrm{m}}|/|tg|$ i.e. reduces with the mode size reduction.
To keep the propagation length acceptable for practical implementation
at fixed degree of localization
one should minimize loss tangent $tg$.
Thus, a prospecting like \cite{Boltasseva_2008_Metamaterials}
is needed to propose the optimal choice of materials for our approach (\ref{our_scheme}).

\section{Semi-analytical description}

\begin{figure}
 \center{\includegraphics[scale=0.4]{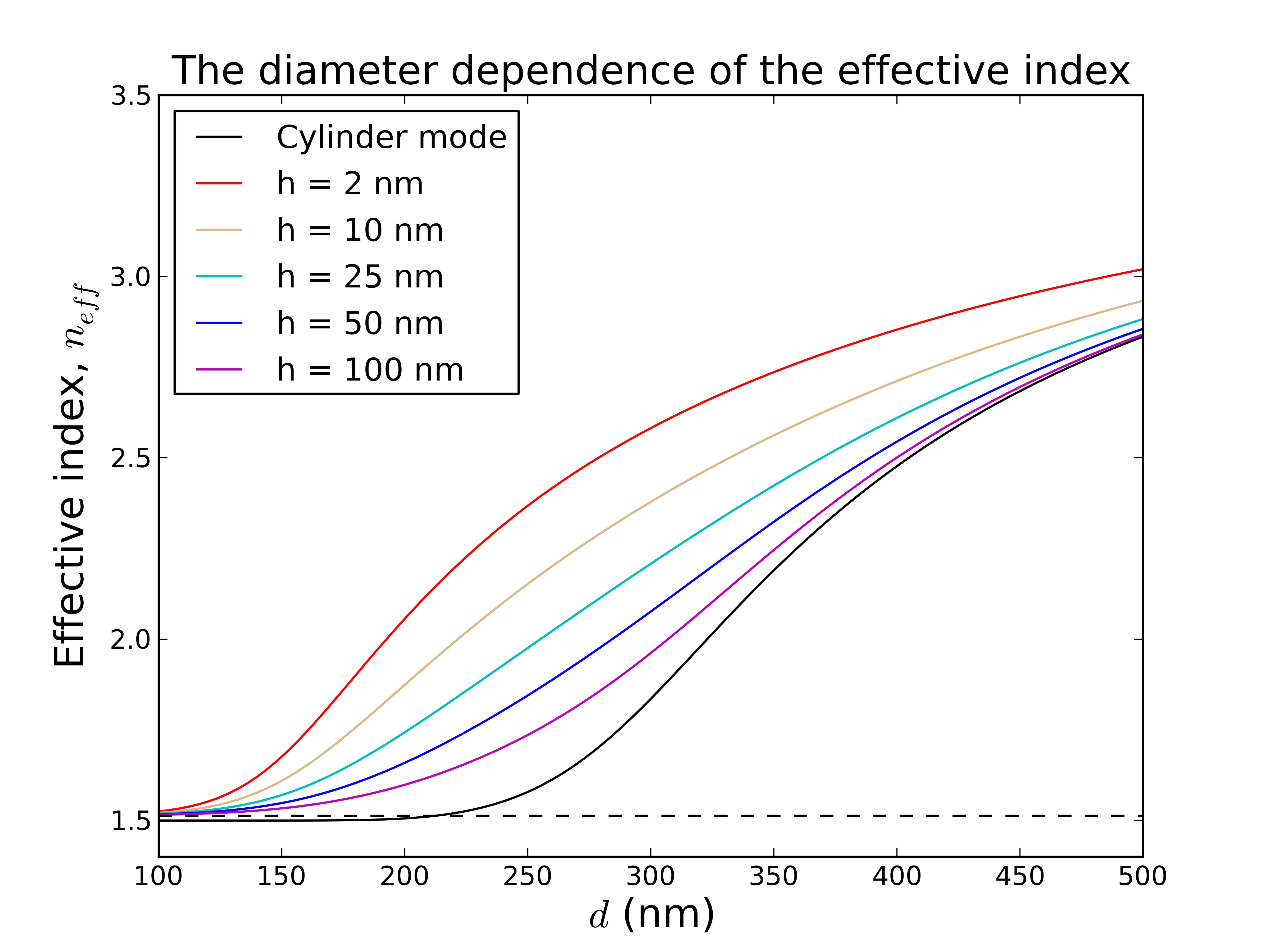}}
  \caption{ (Color online) Effective refractive index of the fundamental hybrid mode versus cylinder
diameter $d$ (coloured lines) compared with those of single
fiber(black solid line) and SPP mode(lower black broken line).
 The dielectric constants  of the cylinder, dielectric and metal are
 $\varepsilon_\mathrm{d}=12.25$ $\varepsilon_\mathrm{g}=2.25$ and
 $\varepsilon_\mathrm{m}=-129+3.3i$ respectively at wavelength $\lambda = 1.55\mu m$.
These parameters are chosen in accordance with the paper \cite{XiangZhang_2008_NaturePhot}.
The critical gap width $h_\mathrm{c} = 5nm$. The HPP-to-CGD crossover point:
$d^\ast\approx 17\mu m$ for $h = 2nm$.}
  \label{pic:comp}
  \end{figure}
In the section, we present the semi-analytical approach to
the propagation of the HPP-mode and discuss the numerical results.
It follows from Maxwell's equations that the electromagnetic
field of guiding mode can be fully described by $z$-components
of the electric and the magnetic fields, $E_z$ and $B_z$
\cite{Marcuse_1972_eng}. These fields satisfy the following
two-dimensional Helmholtz differential equation inside the
homogeneous areas where permittivity is constant:
\begin{equation}
\label{HelmE}
	\Delta^{\!\perp}
	\left\{\!\begin{array}{c}E_z \\ H_z\end{array}\!\right\}
	-
	\left(\beta^2-\varepsilon k^2\right)
	\left\{\!\begin{array}{c}E_z \\ H_z\end{array}\!\right\}
	=0,
\end{equation}
where $\Delta^{\!\perp}=\partial_x^2+\partial_y^2$ and $k=\omega/c$
is the free space wavenumber.
The boundary conditions on the both interfaces are continuity of components $E_z$,
$H_z$, $\varepsilon E_\xi$ and $H_\xi$, where $\xi$-component of a vector
is its normal component.

Our semi-analytical method is based on the
representation of the hybrid waveguide as an integration of the
dielectric fiber and plane plasmonic waveguide. We express the
electromagnetic field of the HPP-mode as a linear combination of cylindrical modes around the
fiber and evanescent plane waves above the metal screen. Boundary conditions
provide the system of linear equations on the expansion
coefficients. Such an approach leads to highly efficient method of numerical solving a difficult
boundary-value problems that describe the propagation of waves in a
complex systems \cite{Bulushev_1988}-\cite{Zakowicz_1997_JOSAA}.  The scheme is developed in detail in Appendix
\ref{section:appendixA}.

To verify our semi-analytical method, in Fig.~\ref{pic:comp} we
present the dependence of the effective index of the fundamental
hybrid mode on the cylinder diameter $d$ for a range of the
gap widths $h$ in the case of telecommunication wavelength when
$\varepsilon_\mathrm{g}<\varepsilon_\mathrm{d}<|\varepsilon_\mathrm{m}|$.
These dispersion curves are obtained from our numerical procedure and
show a good agreement with the results obtained in
\cite{XiangZhang_2008_NaturePhot} by using finite-element package
FEMLab from COMSOL.

\begin{figure}[t]
 \center{\includegraphics[scale=0.4]{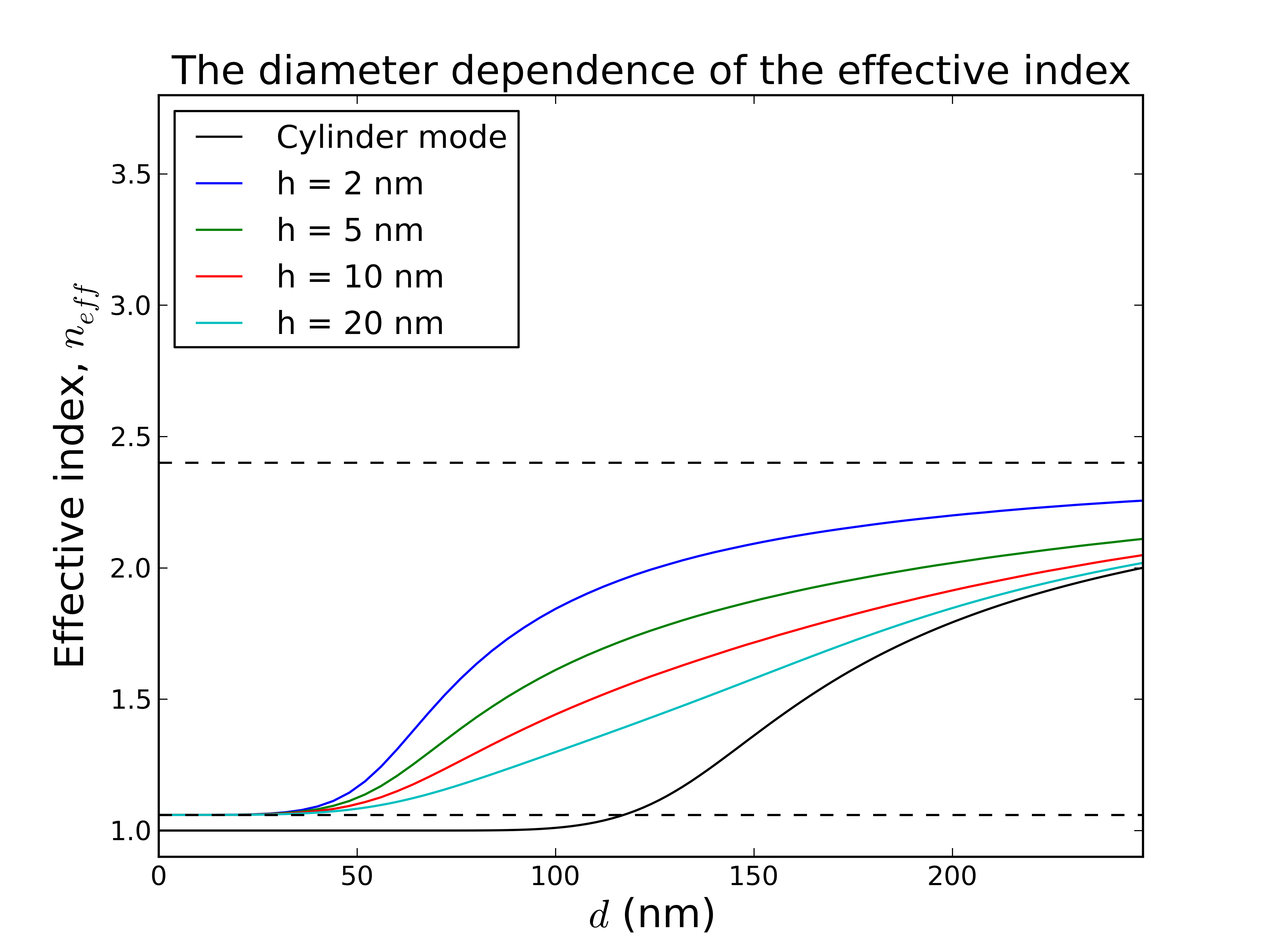}}
  \caption{\small \small Wave vector of the fundamental hybrid mode versus cylinder
diameter $d$ (coloured lines) compared with those of single
fiber(black solid line) and SPP mode(lower black broken line).
 The dielectric constants  of the cylinder, dielectric and metal are
 $\varepsilon_\mathrm{d}=5.76$, $\varepsilon_\mathrm{g}=1$ and
 $\varepsilon_\mathrm{m}=-9.2$ respectively at wavelength $\lambda=0.49 \mu m$.
These parameters are chosen in accordance with the paper \cite{Oulton2009}. The critical gap width
$h_\mathrm{c} = 7nm$. The HPP-to-CGD crossover points are: $d^\ast\approx 310nm$ for $h = 2nm$ , $d^\ast\approx 875nm$ for $h = 5nm$.}
  \label{pic:zhan}
  \end{figure}

In accordance with general argumentation given in
Section~\ref{section:qualitative} we next present two sets of
plots. 
Fig.~\ref{pic:zhan}$(a)$ corresponds to the case of fiber with comparatively low refractive
index, $\varepsilon_{\mathrm{d}}<|\varepsilon_{\mathrm{m}}|$,
the parameters of the waveguide are taken accordingly
to experimental work \cite{Oulton2009}.
Fig.~\ref{pic:zhan}$(b)$ corresponds to opposite limit, when
$\varepsilon_{\mathrm{d}}>|\varepsilon_{\mathrm{m}}|$.
Parameters of these two plots differ only for
metal permittivity $\varepsilon_{\mathrm{m}}$,
the value $\varepsilon_{\mathrm{m}}=-4$ is chosen for Fig.~\ref{pic:zhan}$(b)$.
Here, we do not concretize the material of the metal screen,
our goal is just to demonstrate the qualitative
difference of the guiding mode properties for the case (\ref{our_scheme}).

Presented results indicate that when fiber diameter $d$ is decreased,
the HPP-mode loses confinement along the metal and eventually (at $d
= 0$) becomes a surface plasmon-polariton mode of the flat metal-vacuum
interface. Herewith the effective index of the HPP-mode
monotonically decreases to that of this SPP-mode. Thus all dispersion
curves have the same asymptotic $n_{\mathrm{eff}}\to
n_{\mathrm{mg}}=\sqrt{\varepsilon_\mathrm{m}\varepsilon_\mathrm{g}/(\varepsilon_\mathrm{m}+\varepsilon_\mathrm{g})}$ at small $d$.
Two different behavior are
possible at the opposite limit of large diameter.
As the diameter $d\to \infty$, the HPP-mode can asymptotically tend either fundamental single fiber
mode or the fundamental mode of the planar three-layer system,
the choice depends on the gap width $h$.
If the gap thickness $h$ is below than $h_\mathrm{c}$ (Eq.(\ref{h_c}))
the HPP-mode approaches the CGD-mode with the diameter increasing.
In the case the crossover between the asymptotics occurs at $d^\ast$(black arrows at Fig.~\ref{pic:zhan}$(b)$)
which is determined as
\begin{equation}
\label{crossover}
d^\ast\approx\frac{1}{4(n_{\scriptscriptstyle \mathrm{CGD}}^2-\varepsilon_\mathrm{d})hk^2}.
\end{equation}


For $h > h_\mathrm{c}$ the HPP-mode becomes the cylinder-like in the
limit of the large diameter. In the case the critical diameter $d_0$
corresponding to the transition between small-diameter and
large-diameter asymptotics is defined by the equation
$n_{\scriptscriptstyle \mathrm{SF}}(d_0)=n_{\mathrm{mg}}$, where
$n_{\scriptscriptstyle \mathrm{SF}}(d)$ is the diameter dependence
of the effective index of the single fiber fundamental mode. If the
condition $\sqrt{\varepsilon_\mathrm{d}}kd\ll 1$ is valid one can derive that the localization of this mode is
exponentially small, $n_{\scriptscriptstyle \mathrm{SF}}=
\sqrt{\varepsilon_\mathrm{g}} + \kappa_{g}^2/(2\sqrt{\varepsilon_\mathrm{g}}k^2)$, where
\begin{equation}
\label{single_fiber}
\kappa_\mathrm{g}^2/k^2
\approx
\frac{16e^{-2\gamma+1}}{(kd)^2}
\exp\left\{-\frac{8(\varepsilon_{\mathrm{d}}+\varepsilon_\mathrm{g})}
{\varepsilon_\mathrm{g}(\varepsilon_{\mathrm{d}}-\varepsilon_\mathrm{g})(kd)^2}
\right\}\ll1,
\end{equation}
and $\gamma=0.5772...$  is Euler-Mascheroni constant.

\begin{figure}[t]
 \center{\includegraphics[scale=0.4]{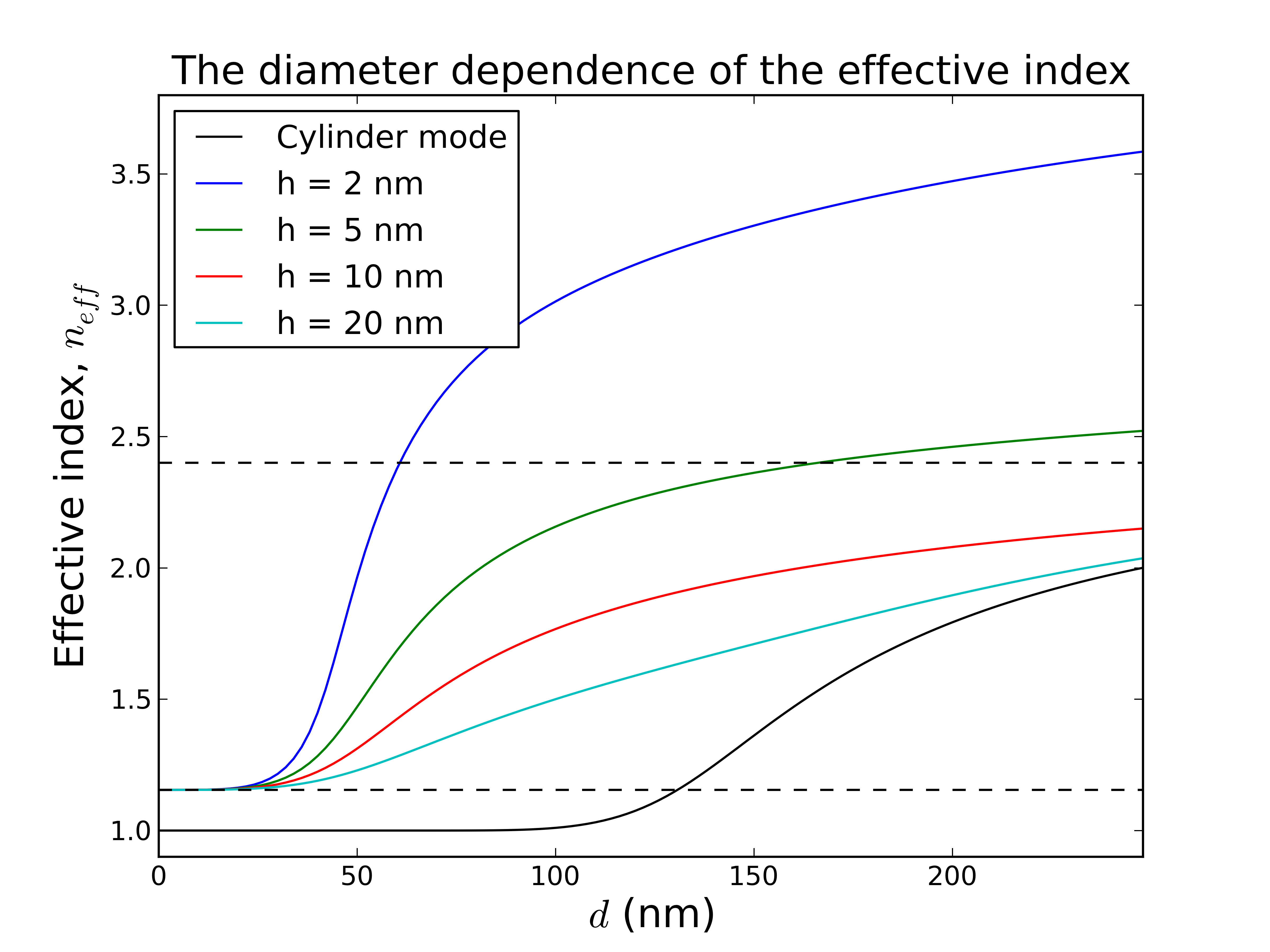}}
  \caption{\small \small Wave vector of the fundamental hybrid mode versus cylinder
diameter $d$(coloured lines) compared with those of single
fiber(black solid line) and SPP mode(lower black broken line).
 The dielectric constants  of the cylinder, dielectric and metal are
 $\varepsilon_\mathrm{d}=5.76$, $\varepsilon_\mathrm{g}=1$ and
 $\varepsilon_\mathrm{m}=-4$ respectively at wavelength $\lambda=0.49\mu m$. The critical gap width
$h_\mathrm{c} = 13,4nm$. The HPP-to-CGD crossover points(black arrows) are: $d^\ast\approx 40nm$ for $h = 2nm$, $d^\ast\approx 65nm$ for $h = 5nm$, $d^\ast\approx 220nm$ for $h = 10nm$.}
  \label{pic:anti}
  \end{figure}

Let us suppose the effective index $n_{\scriptscriptstyle
\mathrm{CGD}}$ of the CGD-mode not to be significantly above that of
bulk plane wave in the fiber medium, $\sqrt{\varepsilon_d}$.
For example, this assumption is true at the conventional plasmonic condition when the absolute value of metal
permittivity is sufficiently greater than the dielectric permittivity,
$|\varepsilon_{\mathrm{m}}|/\varepsilon_{\mathrm{d}}-1\gtrsim 1$.
Just the case is realised at Fig.\ref{pic:comp} and Fig.\ref{pic:zhan}$(a)$.
Then the field penetration depth into the upper dielectric is quite large
as well as the HPP-to-CGD crossover diameter
$d^\ast\gtrsim\lambda$, so the CGD-mode does not provide the strong confinement.
Therefore there are no advantages of CGD-like limit in the case from the view of HPP-mode confinement.
For a given frequency and gap width the choice with the strongest coupling
of the fiber mode and the surface plasmon polariton mode, corresponding to $d=d_0$,
provides the strongest localization of the
field within nanogap due to the great contrast of permittivities
\cite{XiangZhang_2008_NaturePhot}.
At the same time significant part of energy is transferred inside the fiber,
thus the waveguide mode confinement is achieved largely due
to the boundedness of the high-permittivity dielectric part of the waveguide.
Once the diameter of the fiber is optimum and the gap width is
small enough the advantages of the hybrid architecture are used
completely: cross section size of the system can be much less than
the wavelength and mode confinement is much stronger than for uncoupled single
fiber or flat metal-dielectric interface.
To achieve further increase of the HPP-mode confinement the fiber with higher dielectric
constant should be used.

Next let us assume that effective index of the CGD-mode is significantly
larger than the refractive index of the fiber medium.
This can be achieved by diminishing the gap thickness in the case $|\varepsilon_m|<\varepsilon_d$ that corresponds to the Fig.\ref{pic:zhan}$(b)$.
Then the CGD-mode has strong confinement so the crossover diameter can be
decreased to deep subwavelength scale, $d^\ast\ll\lambda$, by tuning the gap width.
Therefore the attractive CGD-like asymptotic can be achieved by HPP-mode
with very small diameter of cylinder providing the wished structure of the
mode with the strong transversal localization in two dimensions within
the gap region and exponential decaying into the cylinder.
Note that in the case the top part of the fiber cross section
is at distances much larger than $1/\kappa_{\mathrm{d}}$ from the gap and its particular shape 
does not play role any more.

\section{Conclusion}

In the paper we have proposed the novel approach for hybrid plasmonic waveguide design providing wide opportunities for
HPP-mode property controlling.
When the absolute permittivity of the metal is less than that of the dielectric the hybrid effective index is
unlimitedly diverges (Eq.(\ref{n_CGD})) with gap width decreasing.
High effective index provides strong confinement of the electromagnetic field in two dimensions within the nanometer-scale gap region.
Thus the mode size can be simply controlled by tuning the waveguide' geometry at fixed frequency and materials constituting the
waveguide.
The advantages of the case $|\varepsilon_m|<\varepsilon_d$ are confirmed
by both qualitative analysis within planar three-layer model and
rigorous semi-analytical method describing the HPP-mode propagation
in general.
The propagation distance of hybrid mode reduces with the mode size reduction.
To achieve long-range propagation at fixed degree of localization one should minimize loss tangent $tg=|\varepsilon_m''/\varepsilon_m'|$ of the metal.
It should be noted that simultaneous satisfying of both conditions $|\varepsilon_m|<\varepsilon_d$ and $tg\ll1$ at optical and near infrared frequencies  is a challenging task.
Thus implementation of the waveguide loss compensation techniques would be required to use such hybrid waveguide as a component of the miniaturized photonic circuits.
Another potential application of our waveguide design lies in study field of the resonant plasmons.
The resonance condition for a surface plasmon-polariton at a planar metal-dielectric interface is the fine-tuning of the permittivities, $-(\varepsilon_m+\varepsilon_d)\ll \varepsilon_d$ \cite{Avrut_2004}.
Thus for particular metal the resonance can be achieved only in a narrow spectral range.
While resonant increasing of the CGD-mode effective index requires only the geometrical condition $h\ll\lambda$.
Thus for the conductor-gap-dielectric structure the resonance of plasmonic mode can be attained by gap width decreasing at any frequency as long as the condition $|\varepsilon_m|<\varepsilon_d$ is valid.

\appendix
\section{Numerical method}
\label{section:appendixA}

The theoretical description of the hybrid waveguide is inhibited by
its complex geometry. In general we should chose such system of
coordinates where the surfaces of the waveguide are the isolines and
Helmholtz equation can be solved by separation of variables. The
hybrid geometry corresponds to the so called bipolar coordinates
based on two sets of orthogonal circles. In this coordinate system
the Helmholtz equation has quite complicated form and accordingly
the set of eigen functions cannot be found analytically. However the
unknown hybrid eigen functions can be expressed in terms of known
solutions of the Helmholtz equation in other coordinate systems.
It is convenient to represent the total electromagnetic field of
HPP-mode as the superposition of the all modes of single
fiber(cylindrical functions) and all SPP modes(evanescent plane
waves) with some unknown coefficients of expansion.

\begin{figure}
 \center{\includegraphics[scale=.85]{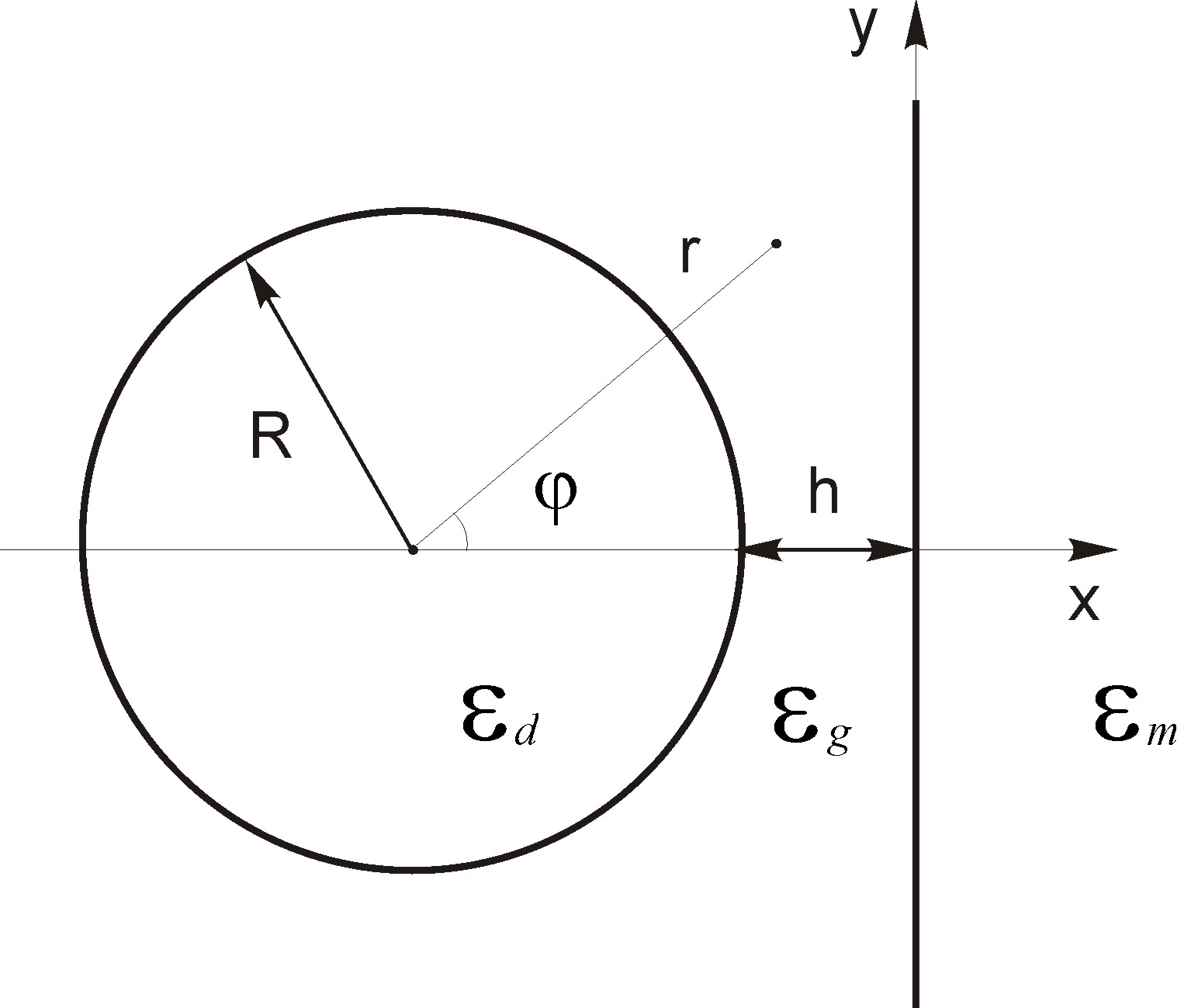}}
  \label{pic:num}
  \caption{Reference system.}
  \end{figure}

We chose the Cartesian system of coordinates with origin at axis of the cylinder which is $z$-axis (Fig.~\ref{pic:num}). 
Supposing the structure of the fundamental hybrid mode to be
symmetric with respect to $x$-axis we can describe the longitudinal
component of the electric field as

\begin{equation}
\label{E}
\left\{ \begin{array}{ll}
E_z^{(\mathrm{d})}=\sum_{n=0}^{\infty}a_n^{{\rm E}}J_n(\chi_\mathrm{d}r)\cos n\varphi\\
E_z^{(\mathrm{g})}=\sum_{n=0}^{\infty}b_n^{{\rm E}}K_n(\kappa_\mathrm{g}r)\cos n\varphi+\\
+\int_{0}^{\infty}c^{{\rm E}}_q\exp(Q\kappa_\mathrm{g}(x-D))\cos q\kappa_\mathrm{g}y\ {\rm d}q\\
E_z^{(\mathrm{m})}=\int_{0}^{\infty}d^{{\rm E}}_q\exp(-Q\kappa_{\mathrm{m}}(x-D))\cos q\kappa_{\mathrm{m}}y\ {\rm d}q\\
\end{array} \right.
\end{equation}
and $z$-components of the magnetic field as
\begin{equation}
\label{H}
\left\{ \begin{array}{ll}
H_z^{(\mathrm{d})}=\sum_{n=0}^{\infty}a_n^{{\rm H}}J_n(\chi_\mathrm{d}r)\sin n\varphi\\
H_z^{(\mathrm{g})}=\sum_{n=0}^{\infty}b_n^{{\rm H}}K_n(\kappa_\mathrm{g}r)\sin n\varphi+\\
+\int_{0}^{\infty}c^{{\rm H}}_q
\exp(Q\kappa_\mathrm{g}(x-D))\sin q\kappa_\mathrm{g}y\ {\rm d}q\\
H_z^{(\mathrm{m})}=\int_{0}^{\infty}d^{{\rm H}}_q\exp(-Q\kappa_{\mathrm{m}}(x-D))\sin
q\kappa_{\mathrm{m}}y\ {\rm d}q
\end{array} \right.
\end{equation}
where $Q=\sqrt{1+q^2}$ and $D=d/2+h$.

To write the corresponding equations, it is convenient to express
the field inside dielectric in terms of only plane evanescent
waves, when we impose the continuity conditions on the boundary of
the metal, and in terms of angular harmonics, for the cylindrical
surface. We solve it by using the evanescent plane wave expansion of
modified cylindrical functions and angular harmonic spectrum of the
evanescent plane waves \cite{Mac}

\begin{equation}
K_n(\kappa_\mathrm{g}r)\cos
n\varphi=\int_{0}^{\infty}F_n^{{\rm E}}(q)e^{-Q\kappa_\mathrm{g}x}\cos
q\kappa_\mathrm{g}y\ {\rm d}q,
\end{equation}
\begin{equation}
K_n(\kappa_\mathrm{g}r)\sin
n\varphi=\int_{0}^{\infty}F_n^{{\rm H}}(q)e^{-Q\kappa_\mathrm{g}x}\sin
q\kappa_\mathrm{g}y\ {\rm d}q,
\end{equation}

\begin{eqnarray}
e^{Q\kappa_\mathrm{g}x}\cos q\kappa_\mathrm{g}y=\sum_{n=0}^{\infty}G_n^{{\rm E}}\cos n\varphi,\\
e^{Q\kappa_\mathrm{g}x}\sin
q\kappa_\mathrm{g}y=\sum_{n=0}^{\infty}G_n^{{\rm H}}\sin n\varphi,
\end{eqnarray}
where

\begin{eqnarray}
F_n^{{\rm E}}=\frac{(Q+q)^n+(Q-q)^n}{2Q},\\
F_n^{{\rm H}}=\frac{(Q+q)^n-(Q-q)^n}{2Q},
\end{eqnarray}

\begin{eqnarray}
G_n^{{\rm E}}&=&\frac{2-\delta_{0n}}2((Q+q)^n+(Q-q)^n)I_n(\kappa_\mathrm{g}r),
\ \ \ \ \ \ \\ 
G_n^{{\rm H}}&=&((Q+q)^n-(Q-q)^n)I_n(\kappa_\mathrm{g}
r).
\end{eqnarray}

Thus the electromagnetic fields in surrounding medium close to the dielectric
waveguide can be written as
\begin{eqnarray}
 \nonumber & \displaystyle\hskip-100pt
    E_z^{(\mathrm{g})}=\sum_{n=0}^{\infty}b_n^{{\rm E}}K_n(\kappa_\mathrm{g}r)\cos n\varphi\ +
    \\[5pt]&\nonumber \displaystyle \ \hskip40pt + \
  \sum_{n=0}^{\infty}\cos n\varphi\int_{0}^{\infty}c^{{\rm E}}_qG_n^{{\rm E}}(r,q)e^{-Q\kappa_\mathrm{g}D}\ {\rm d}q,
\end{eqnarray}
\begin{eqnarray}
 \nonumber & \displaystyle\hskip-100pt
    H_z^{(\mathrm{g})}=\sum_{n=0}^{\infty}b_n^{{\rm H}}K_n(\kappa_\mathrm{g}r)\sin n\varphi\ +
    \\[5pt]&\nonumber \displaystyle \ \hskip40pt + \
  \sum_{n=0}^{\infty}\sin n\varphi\int_{0}^{\infty}c^{{\rm H}}_qG_n^{{\rm H}}(r,q)e^{-Q\kappa_\mathrm{g}D}\ {\rm d}q.
\end{eqnarray}
The corresponding expressions for the fields close to the surface of the metal are
\begin{eqnarray}
 \nonumber & \displaystyle\hskip-30pt
   E_z^{(\mathrm{{\rm g}})}=\int_{0}^{\infty}\sum_{n=0}^{\infty}b_n^{{\rm E}}F_n^{{\rm E}}(q)e^{-Q\kappa_\mathrm{g}x}\cos q\kappa_\mathrm{g}y\ {\rm d}q\ +
    \\[5pt]&\nonumber \displaystyle \ \hskip40pt + \
  \int_{0}^{\infty}c^{{\rm E}}_qe^{Q\kappa_\mathrm{g}(x-D)}\cos q\kappa_\mathrm{g}y\ {\rm d}q,
 \end{eqnarray}

 \begin{eqnarray}
 \nonumber & \displaystyle\hskip-30pt
   H_z^{(\mathrm{g})}=\int_{0}^{\infty}\sum_{n=0}^{\infty}b_n^{{\rm H}}F_n^{{\rm H}}(q)e^{-Q\kappa_\mathrm{g}x}\sin q\kappa_\mathrm{g}y\ {\rm d}q\ +
    \\[5pt]&\nonumber \displaystyle \ \hskip40pt + \
 \int_{0}^{\infty}c^{{\rm H}}_q
e^{Q\kappa_\mathrm{g}(x-D)}\sin q\kappa_\mathrm{g}y\
{\rm d}q.
 \end{eqnarray}

It can be easily derived from the Maxwell's equations that for the normal components
\begin{eqnarray}\label{boundary_conditions}
	E_{\xi}
	& =&
	-
	\frac{i\beta}{\beta^2-\varepsilon k^2}\frac{\partial E_z}{\partial\xi}
	+
	\frac{i k   }{\beta^2-\varepsilon k^2}\frac{\partial H_z}{\partial\eta},\\
	H_{\xi}
	&=&
	-
	\varepsilon\frac{i k   }{\beta^2-\varepsilon k^2}\frac{\partial E_z}{\partial\eta}
	-
	\frac{i\beta}{\beta^2-\varepsilon k^2}\frac{\partial H_z}{\partial\xi},
\end{eqnarray}
where $\eta$ is tangent to the interface coordinate in the transversal plane.

The continuity conditions on the metal surface for $E_z$, $B_z$, $\varepsilon E_x$ and $\varepsilon B_x$
lead to the first system of linear homogeneous equations (SLE) on coefficients
$b_n^{{\rm E}}$, $b_n^{{\rm H}}$, $c_q^{{\rm E}}$, $c_q^{{\rm H}}$, $d_p^{{\rm E}}$, $d_p^{{\rm H}}$.
The corresponding continuity conditions for $E_z$, $B_z$,
$\varepsilon E_r$ and $\varepsilon B_r$ on the cylindrical surface produce
the second SLE on amplitudes $a_n^{{\rm E}}$, $a_n^{{\rm H}}$, $b_n^{{\rm E}}$, $b_n^{{\rm H}}$, $c_q^{{\rm E}}$, $c_q^{{\rm H}}$,
which is now integral with respect to $c_q^{{\rm E}}$, $c_q^{{\rm H}}$.
In order to avoid integration of the unknown functions we express the coefficients $c_q^{{\rm E}}$, $c_q^{{\rm H}}$ in the terms of $b_n^{{\rm E}}$, $b_n^{{\rm H}}$
from the first SLE and substitute them into the second SLE.
The procedure leads to the
infinite system of linear homogeneous algebraic equations for
coefficients $a_n^{{\rm E}}$, $a_n^{{\rm H}}$, $b_n^{{\rm E}}$, $b_n^{{\rm H}}$.
In order to solve the system numerically 
one should truncate it to a finite size. 
Then the propagation constant of the fundamental hybrid mode can be determined from the condition of vanishing of the characteristic determinant.

\section*{Acknowledgments}

We thank I. R. Gabitov for valuable advices and V. V. Lebedev
for fruitful discussions. The work was partially supported by the Russian Federation Government FRBR 12-02-01365-a,
 Council of the President of the Russian Federation grant No. ÑÏ-324.2012.5, the Russian FTP "Kadry", and foundation Dynasty.

\end{document}